\documentclass[twocolumn,english,aps,notitlepage,superscriptaddress]{revtex4-1}
\usepackage{amsmath}
\usepackage{newtxmath}
\usepackage[T1]{fontenc}
\usepackage[latin9]{inputenc}
\usepackage{dsfont}
\usepackage{graphicx}
\usepackage{geometry}
\usepackage{xcolor} 
\usepackage{ulem}
\makeatletter

\newcommand*\LyXThinSpace{\,\hspace{0pt}}

\makeatother

\usepackage{babel}
\begin{document}
\title{Reinterpretation of chiral anomaly on a lattice}
\author{Huan-Wen Wang}
\address{School of Physics, University of Electronic Science and Technology
of China, Chengdu 611731, China}
\author{Bo Fu}
\email{fubo@gbu.edu.cn}

\address{School of Sciences, Great Bay University, Dongguan 523000, China}
\author{Shun-Qing Shen}
\email{sshen@hku.hk}

\address{Department of Physics, The University of Hong Kong, Pokfulam Road,
Hong Kong, China}
\begin{abstract}
The chiral anomaly is a quantum mechanical effect for massless Dirac fermions in both particle physics and condensed matter physics. Here we present a set of effective models for single massless Dirac fermions in one- and three-dimensions in the whole Brillouin zone from higher-dimensional Chern insulators, which uniquely capture both the chiral fermion behavior near the Dirac point and high-energy states at Brillouin zone boundaries. In the presence of electromagnetic fields, the chiral coefficient $C_{5}$ is found to be chemical-potential dependent in general, but quantized precisely at the Fermi surface where chiral symmetry is preserved. This result provides an alternative interpretation of chiral anomaly on a lattice: the anomaly is caused by the symmetry-broken states far below the Fermi surface, and protected by the local chiral symmetry. Our analysis here might provide a potential theoretical foundation for applying the concept of chiral anomaly in condensed matter physics.
\end{abstract}
\maketitle

\section{Introduction}

Chiral anomaly is a fascinating phenomenon that arises in quantum
field theory and has significant implications in various branches
of physics, including particle physics and condensed matter physics
\citep{adler1969,Bell1969,fujikawa1979,peskin_book,armitage2018rmp,lu-FP-2017,Nielsen_plb_1983,Azee-book,Stephanov-prl-2012}.
At its core, chiral anomaly arises from the interplay between chiral
fermions and gauge fields. In the presence of external electromagnetic
fields, the chiral anomaly leads to an imbalance between the rates
at which left-handed and right-handed fermions are generated or annihilated
\citep{Nielsen_plb_1983,DTSon_prb_2013,Stephanov-prl-2012,peskin_book,armitage2018rmp,lu-FP-2017,Azee-book}.
This intriguing effect has profound consequences in various physical
systems. In condensed matter physics, the chiral anomaly has attracted
substantial attentions, particularly in topological materials like
Weyl semimetals and Dirac semimetals, where it gives rise to unique
transport phenomena \citep{DTSon_prb_2013,Burkov_prl_2014,huang-prb-2017,armitage2018rmp,kim-prl-2013,xiong-science-2015,zhang-nc-2016,Li-nc-2016,Liang-prx-2018,Li16natphys,lu-FP-2017}.
Although the relation between chiral anomaly effect and regularization
techniques has been widely discussed in quantum field theory \citep{Wilson1977book,jackiw1969pr,Rothe1998book,Fujikawa-prd-1980,Nielsen1981PLB,Neuberger-prl-1998,Kaplan-plb-1992,Kaplan-prl-2024,Kaplan-prl-2024-1},
it has rarely been explored in condensed matter physics.

Over the past several decades, field theorists have developed diverse
approaches to understanding the chiral anomaly, each offering unique
insights into its origins. In standard perturbation theory, the anomaly
arises from the inability to simultaneously satisfy the Ward identities
for both vector and axial currents in the one-loop triangle diagram
\citep{adler1969,Azee-book}. Fujikawa\textquoteright s functional
integral approach attributes the anomaly to the non-invariance of
the path-integral measure under chiral transformations, where the
Jacobian factor introduces an anomalous term \citep{fujikawa1979,Fujikawa-prd-1980}.
Alternatively, one can define the chiral current operator by placing
the two fermion fields at distinct points separated by a distance
$\epsilon$, and the singularity produced by taking the limit as the
two fields approach each other \citep{jackiw1969pr,peskin_book}.
Pauli-Villars regularization, meanwhile, introduces massive fermions
to cancel divergences at the cost of explicit chiral symmetry breaking.
All the methods mentioned above are based on a low energy effective
model, meanwhile, the anomaly was intensively studied on a lattice
as well, while it is usually struggling with the fermion doubling
problem for a lattice fermion \citep{Wilson1977book,Nielsen1981PLB}.
Efforts to circumvent this---such as domain wall or overlap fermions---have
linked chiral anomalies to topological systems \citep{Neuberger-prl-1998,Callan-npb-1985}.
Notably, topological band theory predicts massless boundary fermions
in topological materials, analogous to domain wall fermions \citep{Qi2008QFT,Qi2011rmp,Moore2010nature,Hasan2010rmp,SQS}.
In Ref. \citep{Fu-npj-2022}, Fu et al further connected anomaly to
condensed matter systems through quantum anomalous semimetals (QAS).
In two dimensions, the relation between parity symmetry and parity
anomaly has been clarified based on the semi-magnetic topological
insulators \citep{Zou-prb-2022,zou-prb-2023,Wang-prb-2024,Fu-arxiv,Chen-2024-scpma,SQS-HQH,Bai-arxiv-2024,Mogi-np-2022}.
However, the connection between chiral symmetry and chiral anomaly
in one or three dimensional QAS has not been satisfactorily built
\citep{Wang-prb-2022}.

In this work, we derived effective models for the quasi-boundary states
of 2D and 4D Chern insulators in the whole momentum space, encompassing both continuum and lattice
formulations. These states exhibit linearized Dirac-like dispersion
within the bulk band gap, aligning with the idealized framework commonly
discussed in prior literature. However, as momentum scales increase,
the initially localized boundary states transition towards bulk states,
acquiring a Dirac mass in the process, consequently breaking chiral
symmetry explicitly. We then analyze the chiral anomaly associated
with these states. In contrast to the conventional regularization
approach in the field theory, the effective model utilized in this
study captures the chiral fermion behavior near the Dirac point while
simultaneously incorporating symmetry-breaking contributions at high
energies. In the presence of electromagnetic field, the calculated
chiral coefficient $C_{5}$ in the continuity equation of the chiral
current is a function of the chemical potential, and it is quantized
in a finite range. Furthermore, the quantization of $C_{5}$ is proved
to be protected by the local chiral symmetry near the Fermi surface.
This result provides an alternative interpretation of chiral anomaly
on a lattice: the anomaly is caused by the symmetry broken states
far below the Fermi surface, and protected by the local chiral symmetry.

\section{Boundary States of 2D Chern insulator\label{sec:2D-Chern-insulator}}

As demonstrated in Ref. \citep{zou-prb-2023,Bai-arxiv-2024}, the
single Dirac fermion in two dimensions can be realized on the boundary
of three-dimensional topological insulator. Following the same procedure,
we proved that the single Dirac fermion in odd dimensions exists in
the boundary of even dimensional topological systems. In this part,
we derived the continuum model of edge states of two dimensional Chern
insulators (CIs) analytically. Here we begin with the effective $2\times2$
modified Dirac equation \citep{SQS},
\begin{equation}
H_{\mathrm{MD}}=v\hbar\left(k_{x}\sigma_{x}+k_{y}\sigma_{y}\right)+m_{0}(\mathbf{k})\sigma_{z},\label{eq:continuum model}
\end{equation}
where $m_{0}(\mathbf{k})=m_{0}-b(k_{x}^{2}+k_{y}^{2})$ is the momentum-dependent
Dirac mass, $\mathbf{k}=(k_{x},k_{y})$ is the wave vector, $m_{0},v,b$
are the model parameters, without loss of generality, we assume $v>0$.
The Chern number of this model can be found as $\mathcal{C}=-\left[\mathrm{sgn}(m_{0})+\mathrm{sgn}(b)\right]/2$.
For CIs, $\mathcal{C}=\pm1$, which requires $m_{0}b>0$. To obtain
the edge states of CIs, we divide Eq. (\ref{eq:continuum model})
into two parts as $H_{\mathrm{MD}}=H_{1d}+v\hbar k_{x}\sigma_{x}$
with $H_{\mathrm{1d}}=v\hbar k_{y}\sigma_{y}+m_{0}(\mathbf{k})\sigma_{z}$.
For a finite length $L$ along the $y$ direction, we make a substitution
as $k_{y}\to-i\partial_{y}$. To solve the eigen equation $H_{\mathrm{1d}}\Psi=E\Psi$,
we set the trial function as $\Psi=e^{i\xi y}\Phi$. The eigen equation
is reduced to $\left[v\hbar\xi\sigma_{y}+m_{0}\left(k_{x},\xi\right)\sigma_{z}\right]\Phi=E\Phi$
with $m_{0}(k_{x},\xi)\equiv m_{0}-b(k_{x}^{2}+\xi^{2})$. For a nontrivial
solution of $\Phi,$ the secular equation of $\xi$ and $E$ gives
four solutions for $\xi(E)$ in Eq. (\ref{eq:solution_xi}), which
is labeled as $\xi_{\alpha s}=s\xi_{\alpha}$ with $p=\pm1,\alpha=1,2$.
In the limit of $L\to+\infty$ , we obtained the eigenenergies in
Appendix \ref{sec:Solution-of-E} as $E_{s}=sm(k_{x}),$ where $m(k_{x})=b\left(k_{c}^{2}-k_{x}^{2}\right)\Theta\left(k_{x}^{2}-k_{c}^{2}\right)$,
$k_{c}=\sqrt{m_{0}/b}$, $s=\pm$ is the band index, and $\Theta(x)$
is the Heaviside step function indicating that $E_{s}$ is zero within
the bulk band gap.

\begin{figure}
\begin{centering}
\includegraphics[width=8cm]{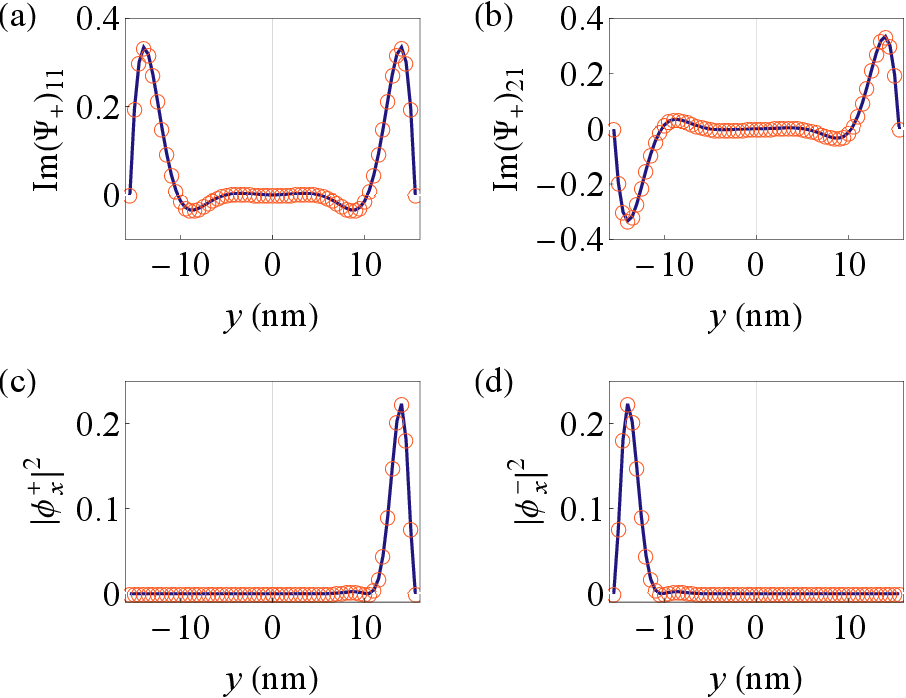}
\par\end{centering}
\caption{\label{fig:(a-b)-Spatial-distributions}(a-b) Spatial distributions
of the imaginary part of $\Psi_{+}(k_{x},y)$ along $y$ direction
for $k_{x}=0.01\,\mathrm{nm^{-1}}$ and $E_{+}=0$. (c-d) Density
distributions ($|\phi_{x}^{\pm}(k_{x},y)|^{2}$) of the two edge states
along $y$ direction for $k_{x}=0.01\,\mathrm{nm^{-1}}$. The orange
circles are calculated from Eq. (\ref{eq:wave_func_1}) numerically,
and the blue solid lines are given by the analytical results in Eq.
(\ref{eq:WF_Edge States}). The length of the open ribbon is chosen
as $L=31\,\mathrm{nm}$. The model parameters are set as $v\hbar=0.44\,\mathrm{eV}\cdot\mathrm{nm}$,
$m_{0}=0.28\,\mathrm{eV}$, $b=0.5\,\mathrm{eV}\cdot\mathrm{nm}^{2}$.}
\end{figure}

For these zero modes, $\mathrm{Im}\xi_{1,2}\ne0$, the corresponding
wave functions can be reduced to the following form \citep{zhou-prl-2008,Bai-arxiv-2024,Fu-nc-2024}
\begin{align}
\Psi_{s} & =\tilde{n}^{\prime}\sin\left[\tilde{\xi}_{0}\left(\frac{L}{2}-|y|\right)\right]\frac{\sigma_{x}+\sigma_{z}}{2}\begin{pmatrix}e^{\xi_{0}y}\\
se^{-\xi_{0}y}
\end{pmatrix},\label{eq:wave_func_1L}
\end{align}
where $\xi_{0}=\frac{v\hbar}{2|b|}$, $\tilde{\xi}_{0}=\xi_{0}\sqrt{\frac{4b(m_{0}-bk_{x}^{2})}{v^{2}\hbar^{2}}-1}$,
$\tilde{n}^{\prime}=\sqrt{\frac{2\xi_{0}(\xi_{0}^{2}+\tilde{\xi}_{0}^{2})}{\tilde{\xi}_{0}\left(\tilde{\xi}_{0}\sinh(\xi_{0}L)-\xi_{0}\sin(\tilde{\xi}_{0}L)\right)}}$.
Then, $\Psi_{s}$ mainly distributes along the two edges, and decay
exponentially to the bulk region. Moreover, as shown in Fig. \ref{fig:(a-b)-Spatial-distributions}
(a-b), for a finite length $L$ along $y$ direction, one can obtain
the complete solutions of Eq. (\ref{eq:secular equation}) numerically
by plugging Eq. (\ref{eq:solution_xi}) into Eq. (\ref{eq:secular equation}),
which results in a transcendental equation of $E_{+}$. We found that
the functional behavior of $\Psi_{+}$ in a finite $L$ can be well
described by $\Psi_{+}$ in Eq. (\ref{eq:wave_func_1L}) {[}solid
blue lines in Fig. \ref{fig:(a-b)-Spatial-distributions} (a-b){]}
when $\xi_{0}L\gg1$. It is noted that all the results of $\Psi_{+}$
can be associated to $\Psi_{-}$ as $\Psi_{-}=\sigma_{x}\Psi_{+}$.

Then, the projection on the lowest two states gives the quasi-edge
Hamiltonian of $H_{\mathrm{MD}}$ as 
\begin{align}
H_{e} & =v\hbar k_{x}\sigma_{x}+m(k_{x})\sigma_{z},\label{eq:edge H 1d}
\end{align}
where the matrix elements of $H_{e}$ are evaluated from $\left(H_{e}\right)_{ss^{\prime}}=\int_{-L/2}^{L/2}\Psi_{s}^{\dagger}(y)H_{\mathrm{MD}}\Psi_{s^{\prime}}(y)dy$.
It should be noted that the so-called quasi-edge Hamiltonian $H_{e}$
is not a pure edge Hamiltonian. As shown in below, the eigen states
of $H_{e}$ will evolve from the edge states to the bulk states with
the increasing of $k_{x}$. In the low energy regime with $k_{c}^{2}>k_{x}^{2}$,
$m(k_{x})=0$, the quasi-edge Hamiltonian satisfies, $H_{e}\phi_{x}^{\pm}=\pm v\hbar k_{x}\phi_{x}^{\pm},$
$\sigma_{x}\phi_{x}^{\pm}=\pm\phi_{x}^{\pm}$, where $\sigma_{x}$
corresponds to the chiral operator in $1+1$ dimensional theory, and
$\phi_{x}^{\pm}$ correspond to the states with positive or negative
chirality, respectively, in the linear model. As the basis of $H_{e}$
is $\{\Psi_{+},\Psi_{-}\}$, the spatial dependence of $\phi_{x}^{\pm}$
can be found as $\phi_{x}^{s}(k_{x},y)=\frac{1}{\sqrt{2}}\left(\sigma_{0}+s\sigma_{x}\right)\Psi_{+}$,
i.e.,
\begin{equation}
\phi_{x}^{s}(k_{x},y)=\frac{\tilde{n}^{\prime}}{\sqrt{2}}\sin\left[\tilde{\xi}_{0}\left(|y|-\frac{L}{2}\right)\right]e^{s\xi_{0}y}\begin{pmatrix}1\\
s
\end{pmatrix}.\label{eq:WF_Edge States}
\end{equation}
Then, as displayed in Fig. \ref{fig:(a-b)-Spatial-distributions}
(a-b), the left-handed and right-handed chiral edge states localized
near $y=L/2$ and $y=-L/2$, respectively. However, in the high energy
regime with $k_{x}^{2}>k_{c}^{2}$, $m(k_{x})=b\left(k_{x}^{2}-k_{c}^{2}\right)$,
the edge states evolve into the bulk states as $\phi_{x}^{s}(k_{x},y)=\cos\frac{\varphi_{s}}{2}\Psi_{+}+s\sin\frac{\varphi_{s}}{2}\Psi_{-}$
with $\cos\varphi_{s}=\frac{sm(k_{x})}{\sqrt{\left(v\hbar k_{x}\right)^{2}+m^{2}(k_{x})}}$,
which distributes in the whole bulk region. Then, the chirality of
massless fermion is no longer a good quantum number anymore, and the
chiral symmetry is broken explicitly.

\begin{figure}
\begin{centering}
\includegraphics[width=7.5cm]{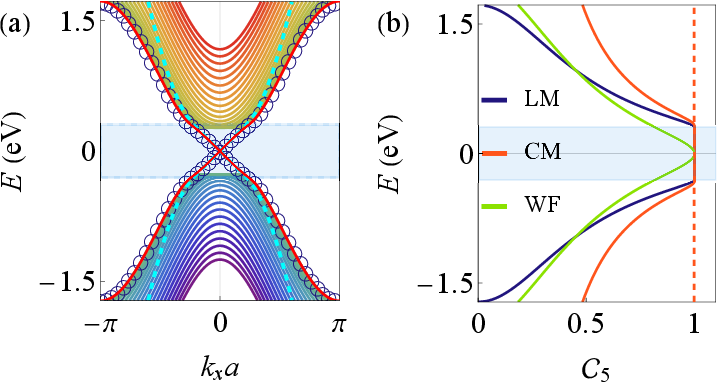}
\par\end{centering}
\caption{\label{fig:C5} (a) Energy spectrum of the Chern insulator on a lattice,
the red lines indicates the two energy bands near the half-filling,
the open circles are evaluated from the lattice model in Eq. (\ref{eq:edge H 1d-1}),
and the dashed line are given by the continuum model in Eq. (\ref{eq:edge H 1d}).
The length of the ribbon is chosen as $L=31\,\mathrm{nm}$ along $y$
direction. (b) The chemical potential-dependence of dimensionless
coefficient $\mathcal{C}_{5}$ for the lattice model (LM), continuum
model (CM) and Wilson fermion (WF). The dashed line indicates the
value of $\mathcal{C}_{5}$ for the case of ideal Dirac fermion with
linear dispersion.The shadow rectangle corresponds to the chiral symmetric
regime with $m(k_{x})=0$.}
\end{figure}

We further study the edge states for a lattice model in Appendix \ref{sec:Lattice-Model},
where we make replacements as $k_{i}\to\frac{1}{a}\sin k_{i}a,$ $k_{i}^{2}\to\frac{2}{a^{2}}\left(1-\cos k_{i}a\right)$
in the continuum Hamiltonian $H_{\mathrm{MD}}$, where $a$ is the
lattice constant. The Chern number for the lattice Hamiltonian is
given by $\mathcal{C}=\left[2\mathrm{sgn}\left(m_{0}-4t_{L}\right)-\mathrm{sgn}\left(m_{0}-8t_{L}\right)-\mathrm{sgn}\left(m_{0}\right)\right]/2$,
where $t_{L}=\frac{b}{a^{2}}$. Without loss of generality, we assume
the band inversion occurs near the $\Gamma$ point, i.e., $0<m_{0}<4t_{L}$
and $\mathcal{C}=-1$, then, the quasi-edge Hamiltonian reads
\begin{align}
H_{e} & =\lambda_{L}\sin k_{x}a\sigma_{x}+m(k_{x})\sigma_{z},\label{eq:edge H 1d-1}
\end{align}
where $m(k_{x})=\tilde{m}(k_{x})\Theta(-\tilde{m}(k_{x}))$, $\tilde{m}(k_{x})=m_{0}-4t_{L}\sin^{2}\frac{k_{x}a}{2}$,
$\lambda_{L}=\frac{v\hbar}{a}$. In Fig. \ref{fig:C5}(a), we numerically
calculate the eigen energies for a ribbon of CIs. The red lines are
the two lowest states near the half-filling, and the open circles
are evaluated from Eq. (\ref{eq:edge H 1d-1}), which has a good agreement
with the numerical calculation in the whole Brillouin zone. Moreover,
we also present the eigen energy of quasi-edge states in Eq. (\ref{eq:edge H 1d}),
which match well with the lattice one within the band gap. As derived
in Appendix \ref{sec:Zero-Modes}, the localized edge states only
exist in a finite momentum region as $\tilde{m}(k_{x})>0$, which
is consistent with the continuum model. In Fig. \ref{fig:(a-b)-Spatial-distributions-S1}(a),
we numerically calculate the edge states in the lattice model, which
can be well depicted by $\phi_{x}^{\pm}(k_{x},y)$ in our continuum
model within the bulk gap. Outside of the critical momentum region,
localized states vanish and bulk states dominate {[}see Fig. \ref{fig:(a-b)-Spatial-distributions-S1}(b){]}.
As discussed below, the bulk states break the chiral symmetry explicitly
and lead to the non-conservation of chiral current in the chiral symmetric
regime.

\section{Chiral anomaly in 1+1 dimensions}

As discussed above, the one-dimensional single massless Dirac fermion
can be realized on the edge of two dimensional CIs. To investigate
the quantum anomaly effect for the one-dimensional state, we rewrite
the effective edge models in Eq. (\ref{eq:edge H 1d}) and Eq. (\ref{eq:edge H 1d-1})
in terms of the gamma matrices as 
\begin{equation}
H_{e}=h_{1}(k_{x})\gamma^{0}\gamma^{1}+m(k_{x})\gamma^{0},\label{eq:he_continuum_lattice}
\end{equation}
where $\gamma^{0}\gamma^{1}=\sigma_{x},\gamma^{0}=\sigma_{z}$,$\gamma^{1}=i\sigma_{y}$,
$h_{1}(k_{x})=v\hbar k_{x}$ for the continuum model and $h_{1}(k_{x})=\lambda_{L}\sin k_{x}a$
for the lattice model. In the chiral symmetry invariant regime ($k_{x}^{2}<k_{c}^{2}$),
$[H_{e},\gamma^{5}]=0$ with $\gamma^{5}=\sigma_{x}$. In the high
energy regime ($k_{x}^{2}>k_{c}^{2}$), the chiral symmetry is explicitly
broken, and $[H_{e},\gamma^{5}]=2m(k_{x})\gamma^{0}\gamma^{5}$. Following
the Jackiw-Johnson approach to the chiral anomaly \citep{jackiw1969pr,Wang-prb-2021},
we can derive the continuity equation for the gauge-invariant chiral
current $j_{5}^{\mu}$ by taking the limit $\epsilon\to0$ as
\begin{equation}
\partial_{\mu}j_{5}^{\mu}=-\lim_{\epsilon\to0}i\frac{e}{\hbar}\epsilon^{\alpha}F_{\mu\alpha}j_{5}^{\mu}(x,\epsilon)+i\bar{\psi}\frac{2m(k_{x})}{\hbar}\gamma^{5}\psi,\label{eq:continuity_1d}
\end{equation}
where $F_{\alpha\mu}=\partial_{\alpha}A_{\mu}-\partial_{\mu}A_{\alpha}$
is the field strength, and $A_{\mu}=(\phi,\mathbf{0})$ with $\phi$
the electric potential. $\psi$ and $\bar{\psi}=\psi^{\dagger}\gamma^{0}$
are the Dirac spinors. The first term on the right-hand side of Eq.
(\ref{eq:continuity_1d}) gives the anomalous correction due to the
quantum fluctuation effect. The second term is tied to the pseudoscalar
condensation and is the consequence of the explicit chiral symmetry
breaking from the Dirac mass $m(k_{x})$. The gauge invariant chiral
currents $j_{5}^{0}$ and $j_{5}^{1}$ are defined as 
\begin{align*}
j_{5}^{0}(x,\epsilon) & =e^{i\phi(x,\epsilon)}\left(\bar{\psi}_{+}\gamma^{0}\gamma^{5}\psi_{-}\right),\\
j_{5}^{1}(x,\epsilon) & =e^{i\phi(x,\epsilon)}\left(\hbar^{-1}\sum_{n}\sum_{p=0}^{n-1}c_{n}(i\partial_{x})^{p}\bar{\psi}\gamma^{1}\gamma^{5}(-i\partial_{x})^{n-p-1}\psi\right),
\end{align*}
where $\psi_{\pm}=\psi(x\pm\frac{\epsilon}{2})$, $\phi(x,\epsilon)=\frac{e}{\hbar}\int_{x-\epsilon/2}^{x+\epsilon/2}A_{\alpha}(x)dx^{\alpha}$,
$c_{n}$ is the coefficient of Taylor expansion of $h_{1}(-i\partial_{x})=\sum_{n}c_{n}(-i\partial_{x})^{n}$.

Now, let us evaluate the anomalous correction from the quantum fluctuation
in Eq. (\ref{eq:continuity_1d}). Here we consider nonzero $\epsilon_{1}$,
the possible nonzero contribution is from $\langle\epsilon^{1}F_{10}j_{5}^{0}\rangle$,
which can be found as 
\begin{align}
\lim_{\epsilon\to0}ie\langle\epsilon^{1}F_{10}j_{5}^{0}\rangle= & \lim_{\epsilon_{1}\to0}eE_{x}\epsilon_{1}\int_{BZ}\frac{dk_{x}}{2\pi}\frac{h_{1}(k_{x})}{\varepsilon(k_{x})}\sin k_{x}\epsilon_{1}\label{eq:quantum_fluctuation}
\end{align}
with $\varepsilon(k_{x})=\sqrt{m^{2}(k_{x})+h_{1}^{2}(k_{x})}$. For
massive Dirac fermion without quadratic momentum correction in the
continuum model, i.e., $b=0$, we can perform the integral in Eq.
(\ref{eq:quantum_fluctuation}) analytically as $\epsilon_{1}\frac{\partial}{\partial\epsilon_{1}}\int_{-\infty}^{+\infty}\frac{dk_{x}}{2\pi}\frac{e^{ik_{x}\epsilon_{1}}}{\sqrt{m_{0}^{2}+v^{2}\hbar^{2}k_{x}^{2}}}=\frac{\epsilon_{1}}{\pi}\frac{\partial}{\partial\epsilon_{1}}K_{0}(|\epsilon_{1}m_{0}|)$,
where $K_{0}(x)$ is the modified Bessel functions of the second kind.
Considering the asymptotic behavior of $K_{0}(x)$ for small argument
$x$, $K_{0}(x)\sim-\gamma-\log\frac{x}{2}$ with $\gamma$ the Euler-Mascheroni
constant, we have $\lim_{\epsilon\to0}ie\langle\epsilon^{1}F_{10}j_{5}^{0}\rangle=\frac{eE_{x}}{\pi}$
\citep{peskin_book}. Then, there is a finite contribution from the
quantum fluctuation in Eq. (\ref{eq:continuity_1d}), which is cancelled
by the vacuum polarization of valance band exactly at the half-filling
\citep{Wang-prb-2021}. When $b\ne0$, $\lim_{\epsilon\to0}ie\langle\epsilon^{1}F_{10}j_{5}^{0}\rangle$
should equal to zero as $\lim_{\epsilon\to0}ie\langle\epsilon^{1}F_{10}j_{5}^{0}\rangle=\frac{eE_{x}}{2\pi}\frac{h_{1}(k_{x})}{\varepsilon(k_{x})}\mid_{-\infty}^{+\infty}=0$.
Then, we only need to calculate the contribution from the explicit
chiral symmetry breaking in Eq. (\ref{eq:continuity_1d}). For the
lattice model, as the Brillouin zone (BZ) is finite, the integral
in Eq. (\ref{eq:quantum_fluctuation}) is finite, the quantum fluctuation
vanishes as $\epsilon_{1}\to0$.

In the presence of electric potential $\phi=eE_{x}x$, we can compute
the expectation value for the Pseudo scalar condensation in the perturbation
theory. First, the perturbed eigenstates can be found as 
\begin{align*}
|s;k_{x}\rangle_{p}= & |s;k_{x}\rangle+i\frac{seE_{x}}{4}\frac{\partial_{k_{x}}\phi_{s}}{\varepsilon(k_{x})}|\bar{s};k_{x}\rangle,
\end{align*}
where $s=\pm$ is the band index, $|s;k_{x}\rangle=[\mathrm{sgn}(h_{1})\cos\frac{\phi_{0s}}{2}|k_{x}\rangle,s\sin\frac{\phi_{0s}}{2}|k_{x}\rangle]^{T}$
is the unperturbed eigenstates with eigen energy $s\varepsilon(k_{x})$
and $\cos\phi_{s}=\frac{sm(k_{x})}{\varepsilon(k_{x})}$. Now, we
can evaluated the expectation value of $im(k_{x})\gamma^{0}\gamma^{5}$
in the perturbed states as $_{p}\langle s;k_{x}|im(k_{x})\gamma^{0}\gamma^{5}|s;k_{x}\rangle_{p}=s\frac{eE_{x}}{2}\frac{d}{dk_{x}}\left(\frac{h_{1}(k_{x})}{\varepsilon(k_{x})}\right)$.
The grand canonical ensemble average of $2im(k_{x})\gamma^{0}\gamma^{5}$
at the zero temperature and chemical potential $\mu$ is evaluated
as
\begin{align}
\langle:\Psi^{\dagger}2im(k_{x})\gamma^{0}\gamma^{5}\Psi:\rangle= & \frac{eE_{x}}{\pi}\frac{h_{1}(k_{x})}{\varepsilon(k_{x})}\bigg|_{k_{\partial}}^{k_{F}},\label{eq:Pseudo-scalar condensation}
\end{align}
where $k_{\partial}$ is the momentum at the boundary of BZ, for instance,
$k_{\partial}=\pi$ for a lattice fermion, and $k_{\partial}=+\infty$
for a continuum model. Alternatively, we can derive the expectation
value of Pseudo-scalar condensation from the linear response theory.
As shown in Appendix \ref{sec:Linear-response-theory}, after a tedious
but straightforward calculation, we recover results from the perturbed
eigenstates in a constant electric field.

In Eq. (\ref{eq:Pseudo-scalar condensation}), when $b\ne0$, the
contribution from the depth of valence band vanishes as $\frac{h_{1}(k_{\partial})}{\varepsilon(k_{\partial})}=0$.
Meanwhile, due to the symmetry breaking term in the high energy, the
anomaly from the gauge-invariant current also vanishes \citep{Wang-prb-2022}.
Then, we can get an anomaly free continuity equation for chiral current
as
\begin{equation}
\partial_{x}j_{x}^{5}+\partial_{t}\rho_{5}=\mathcal{C}_{5}\frac{eE_{x}}{\pi\hbar},\label{eq:continuity_1d-F}
\end{equation}
where $\mathcal{C}_{5}=\left|\frac{h_{1}(k_{F})}{\mu}\right|$ is
the chiral coefficient. When $b=0$ in the continuum model, $\left|\frac{h_{1}(k_{\partial})}{\varepsilon(k_{\partial})}\right|=1$,
which is cancelled by the anomalous correction from the infinite Dirac
sea \citep{Azee-book}. Hence, the above continuity equation is valid
for arbitrary $m_{0}$ and $b$ for a single Dirac fermion.

For the single Dirac fermion described in Eq. (\ref{eq:he_continuum_lattice}),
as shown in Fig. \ref{fig:C5}(b), in the chiral symmetric regime
$(k_{x}^{2}<k_{c}^{2})$, $m(k_{x})=0$ and $\mathcal{C}_{5}=1,$
one reproduces the chiral anomaly equation for an ideal massless Dirac
fermion \citep{peskin_book}, the anomalous term in the right-hand
side of Eq. (\ref{eq:continuity_1d-F}) equation is from the explicit
symmetry breaking term in the high energy. This is very similar with
the half-quantized Hall conductance in two dimensional parity anomalous
semimetals \citep{zou-prb-2023,Wang-prb-2024}. In the high energy
regime, the chiral symmetry is not preserved at the Fermi surface,
the chiral coefficient $\mathcal{C}_{5}$ decreases with the increasing
of fermi energy. At the BZ boundary, the chiral coefficient vanishes
for the lattice model as depicted in Fig. \ref{fig:C5}(b). When $k_{c}\to0$,
Eq. (\ref{eq:he_continuum_lattice}) depicts the Wilson fermion, $\mathcal{C}_{5}$
deviates the quantized value for a finite Fermi energy as the chiral
symmetry is always broken as indicated by the green line in Fig. \ref{fig:C5}(b).

Extending this analysis to the species doubling of Weyl fermions,
we consider a chiral-symmetric lattice Hamiltonian of the form $H=\lambda_{L}\sin(k_{x}a)\gamma^{0}\gamma^{1}$,
which hosts two Dirac points near $\ensuremath{k_{x}=0}$ and $k_{x}=\pi$.
Crucially, the chiral symmetry is preserved across the entire BZ,
and no quantum anomaly arises due to the finite BZ, resulting in the
conservation of the chiral current $(\ensuremath{\mathcal{C}_{5}=0})$.
An intriguing observation emerges even when a constant mass term $m_{0}\gamma^{0}$
is introduced near the Dirac point, yielding $H=\lambda_{L}\sin(k_{x}a)\gamma^{0}\gamma^{1}+m_{0}\gamma^{0}$.
Despite the mass term, the chiral current remains conserved. Here,
two Fermi wavevectors $k_{F1}$ and $\ensuremath{k_{F2}}$ (satisfying
$(k_{F1}+k_{F2})a=\pi$) appear at the Fermi energy, leading to the
vanishing of pseudoscalar condensation: $\langle:\Psi^{\dagger}2im_{0}\gamma^{0}\gamma^{5}\Psi:\rangle=\left.\frac{eE_{x}}{\pi}\frac{\lambda_{L}\sin(k_{x}a)}{\varepsilon(k_{x})}\right|_{k_{F2}}^{k_{F1}}=0.$
This result smoothly connects to the massless limit $(\ensuremath{m\to0})$
of the chiral-symmetric Hamiltonian, demonstrating consistency across
both cases.

\section{Boundary States in 4D Chern insulator}

To extend the above discussion to the three-dimensions, we begin with
a four dimensional Chern insulator, which can be described by the
following $k\cdot p$ model as \citep{Qi2008QFT}
\[
H=\sum_{\mu=1}^{4}v\hbar k_{\mu}\gamma^{0}\gamma^{\mu}+\left(m_{0}-b\sum_{\mu=1}^{4}k_{\mu}^{2}\right)\gamma^{0},
\]
where $\gamma^{0}=\tau_{3},\gamma^{j}=i\tau_{2}\sigma_{j}$ for $j=1,2,3$,
$\gamma^{4}=-i\tau_{1}$ are the gamma matrices. For topologically
nontrivial case, $m_{0}b>0$. To find the solution of boundary state,
we consider the open boundary condition along the fifth-direction
and make a substitution for $k_{4}$ as $k_{4}\to-i\partial_{r_{4}}$
in $H$. Similar to the two-dimensional case, the projection on the
lowest four states gives the quasi-surface Hamiltonian of four-dimensional
Chern insulators as
\begin{align}
H_{\mathrm{surf}}=v\hbar k_{j}\gamma^{0}\gamma^{j}+ & m(k)\gamma^{0},\label{eq:surface_4D}
\end{align}
where $m(k)=\tilde{m}(k)\Theta\left(-\tilde{m}(k)b\right)$ with $\tilde{m}(k)=m_{0}-b\sum_{i=1}^{3}k_{i}^{2}$.
In the chiral symmetry invariant regime ($k<k_{c}=\sqrt{m_{0}/b}$)
$m(k)=0$, and Eq. (\ref{eq:surface_4D}) describes a three-dimensional
gapless Dirac fermion with chiral symmetry. In the high energy regime
with $k>k_{c}$, $m(k)=m_{0}-bk^{2},$ the chiral symmetry of Dirac
fermion is broken explicitly. Similar to the one-dimensional Dirac
fermion, we expect the Dirac fermion in Eq. (\ref{eq:surface_4D})
can exhibit chiral anomaly effect in the presence of electromagnetic
field. Moreover, the three-dimensional Dirac fermion is also expected
to display fractional magneto-electric response \citep{Wang-prb-2022}.

\section{Chiral anomaly in 3+1 dimensions}

As discussed above, the three-dimensional gapless Dirac fermion can
be realized on the boundary of four dimensional Chern insulators.
To study the chiral anomaly effect of the boundary state, we begin
with the effective model in Eq. (\ref{eq:surface_4D}). The chiral
anomaly of a three-dimensional Dirac fermion in a quantized field
can be reduced to the lowest Landau level, which is non-degenerate.

Under a perpendicular magnetic field $\mathbf{B}=B\hat{z}$, the momentum
operators $\hbar k_{i}$ are replaced by kinematic momentum operators
$\Pi_{i}=\hbar k_{i}+eA_{i}$ under the Peierls substitution, where
$A_{i}$ is the $i$-th component of the vector potential. Without
loss of generality, one can choose the vector potential as $A_{i}=-\delta_{i1}By$
with $\delta_{ij}$ the Kronecker delta symbol. By taking advantage
of the ladder operator technique \citep{shen2005prb}, we obtain the
eigenvalues and eigenstates for the operator $\mathbf{\Pi}\cdot\mathbf{\sigma}$,
$\mathbf{\Pi}\cdot\mathbf{\sigma}|nk_{x}k_{z}\chi_{n}\rangle=\chi_{n}\sqrt{\hbar^{2}k_{z}^{2}+n\left(\frac{\Omega}{v}\right)^{2}}|nk_{x}k_{z}\chi_{n}\rangle$,
where $n=0,1,2,\cdots$ are the indices of the Landau levels, and
$\Omega=\sqrt{2}v\hbar/\ell_{B}$ is the cyclotron energy with the
magnetic length $\ell_{B}=\sqrt{\hbar/\left|eB\right|}$. $\chi_{n}=\pm1$
for $n>0$ and $\chi_{0}=-\mathrm{sgn}(Bk_{z})$ for $n=0$ are the
helicity of the operator $\mathbf{\Pi}\cdot\mathbf{\sigma}$. As the
translational symmetry is preserved along the $x$ and $z$ direction,
$k_{x}$ and $k_{z}$ are still good quantum numbers. Then, the energy
spectra of the operator $\mathbf{\Pi}\cdot\mathbf{\sigma}$ are quantized
into a series of Landau bands. The lowest Landau levels are primarily
responsible for the emergence of exotic quantum phenomena, we can
project Eq. (\ref{eq:surface_4D}) onto the lowest Landau levels (LLL).
The Hamiltonian and all the physical quantities can be expressed in
terms of $2\times2$ matrices and the motions of the electrons are
confined along the direction of the magnetic field, i.e.,
\begin{equation}
H_{LLL}=-\mathrm{sgn}(B)v\hbar k_{z}\tau_{1}+m(k_{z})\tau_{3}.\label{eq:LLL}
\end{equation}
Now, we have reduced the three dimensional problem to one dimension.
For the lowest Landau bands, the chiral operator is defined as $\gamma^{5}=-\mathrm{sgn}(B)\tau_{1}$.
If we ignored the Landau degeneracy, Eq. (\ref{eq:LLL}) is equivalent
to Eq. (\ref{eq:edge H 1d}). Consequently, the continuity equation
for the one dimensional chiral current of Eq. (\ref{eq:LLL}) can
be found as $\partial_{z}j_{z}^{5}+\partial_{t}\rho_{5}=\mathcal{C}_{5}\frac{eE_{z}}{\pi\hbar}$.
By further introducing the Landau degeneracy $n_{L}=\frac{|eB|}{2\pi\hbar}$,
one obtain the continuity equation of chiral current for three-dimensional
Dirac fermion as
\begin{equation}
\nabla\cdot\mathbf{J}_{5}+\partial_{t}\rho_{5}=\mathcal{C}_{5}\frac{e^{2}}{2\pi^{2}\hbar^{2}}\mathbf{E}\cdot\mathbf{B}.\label{eq:Chiral equation in 3D}
\end{equation}
Here $\mathcal{C}_{5}$ has the same form as the one in one dimension,
i.e., $\mathcal{C}_{5}=\frac{v\hbar k_{F}}{|\mu|}$. For the three-dimensional
gapless Dirac fermion, $\mathcal{C}_{5}=1$ in the chiral symmetry
invariant regime ($k^{2}<k_{c}^{2}$), one reproduces the chiral anomaly
equation for an ideal massless Dirac fermion in the quantum field
theory \citep{Azee-book,peskin_book}, the anomalous term in the right-hand
side of above equation is from the explicit symmetry breaking term
in the high energy. Actually, we can also consider a lattice structure
along $z-$direction, where we make replacements as $k_{z}\to\frac{1}{a}\sin k_{z}a,$
$k_{z}^{2}\to\frac{2}{a^{2}}\left(1-\cos k_{z}a\right)$ in Eq. (\ref{eq:surface_4D})
and Eq. (\ref{eq:LLL}). Then, we can obtain the continuity equation
of chiral current in Eq. (\ref{eq:Chiral equation in 3D}) with $\mathcal{C}_{5}=\frac{\lambda_{L}\sin k_{z}a}{\sqrt{m^{2}(k_{z})+\lambda_{L}^{2}\sin^{2}k_{z}a}}\mid_{k_{F}}$.

It is worth noting that the continuity equation of chiral current
in odd dimensions relies on the chemical potential or Fermi surface,
as illustrated in Fig. \ref{fig:C5}, which is distinct from the one
given by the quantum field theory. Our analysis here might provide
a potential theoretical foundation for applying the concept of chiral
anomaly in topological materials. Specifically, in the case of Dirac
semimetals, the linear dispersion only exists in a finite momentum
region in the energy spectrum, and there is usually a small band gap
in the Dirac cone, which seems to not satisfy the condition of ideal
Dirac fermions. However, our findings suggest that when the energy
dispersion is linear near the Fermi surface, $\mathcal{C}_{5}\approx1$,
the ideal chiral anomaly equation remains valid. Nevertheless, we
should remind ourself that the non-conservation of chiral current
in the linear dispersion is attributed to the symmetry breaking term
in the high energy, akin to the regularization process in quantum
field theory.

\section{Fractional magneto-electric response}

For the three-dimensional Dirac system, there might be topological
magneto-electric effect, which is characterized by the topological
$\theta$ term \citep{Wilczek1987axiondynamics,Schnyder2008classification,Ryu2010tenfold,Li2010np,Essin2009OMP,Spaldin2005ME,Fiebig2005ME}.
Although the chiral symmetry is broken in high energy in Eq. (\ref{eq:surface_4D}),
the system still possesses an additional sublattice symmetry. For
the general Dirac Hamiltonian $\mathds{h}_{D}=h_{0}\gamma^{0}+\sum_{i=1}^{3}h_{i}\gamma^{0}\gamma^{i}$,
there exists a matrix $\gamma^{0}\gamma^{4}$ which is anti-commutative
with $\mathds{h}_{D}$, i.e., $\{\gamma^{0}\gamma^{4},\mathds{h}_{D}\}=0$,
the topological properties of $\mathds{h}_{D}$ can be described by
the winding number \citep{Schnyder2008classification,Ryu2010tenfold}.
By taking advantage of the properties of gamma matrices, $\{\gamma^{\mu},\gamma^{\nu}\}=2\eta_{\mu\nu}I$
and $\mathrm{tr}[\gamma^{4}\gamma^{\mu}\gamma^{\nu}\gamma^{\sigma}\gamma^{\rho}]=-4\epsilon_{\mu\nu\sigma\rho}$
with ${\displaystyle \eta_{\mu\nu}}$ the Minkowski metric with signature
$(+,-,-,-)$ and $\epsilon_{\mu\nu\sigma\rho}$ the Levi-Civita symbol,
we can obtain the winding number for the Dirac Hamiltonian $\mathds{h}_{D}$
as 
\begin{align}
w_{3D} & =\int d^{3}\mathbf{k}\varpi_{3D},\label{eq:winding}\\
\varpi_{3D} & =\frac{\epsilon_{\mu\nu\sigma\rho}}{2\pi^{2}h^{4}}h_{\mu}\left(\partial_{k_{x}}h_{\nu}\right)\left(\partial_{k_{y}}h_{\sigma}\right)\left(\partial_{k_{z}}h_{\rho}\right),
\end{align}
where the indices $\mu$, $\nu$, $\sigma$, $\kappa$, $\rho$ run
over $0,1,2,3$, $h=\sqrt{\sum_{\beta=0}^{3}h_{\beta}^{2}}$, $\varpi_{3D}$
is the winding number density. Considering the three-dimensional Dirac
fermion in Eq. (\ref{eq:surface_4D}), $h_{0}=m(k),h_{1}=v\hbar k_{x}$,
$h_{2}=v\hbar k_{y}$, $h_{3}=v\hbar k_{z}$, $\{H_{\mathrm{surf}},\gamma^{0}\gamma^{4}\}=0$.
By integrating out Eq. (\ref{eq:winding}), the winding number of
$H_{\mathrm{surf}}$ reads
\begin{align*}
\omega_{3D} & =\frac{1}{2}\mathrm{sgn}(b),
\end{align*}
which is half-quantized and its relation with chiral symmetry near
the Dirac point has been discussed for three-dimensional Wilson fermions
\citep{Wang-prb-2022,Fu-npj-2022}. To explore the magneto-electric
effect related to the half-quantized winding number, we further introduce
a symmetry breaking term $m_{4}\gamma^{0}\gamma^{4}$ in Eq. $\mathds{h}_{D}$
as,
\begin{align}
H_{\mathrm{surf}}^{4}= & \mathds{h}_{D}+m_{4}\gamma^{0}\gamma^{4},\label{eq:surface_4D-1}
\end{align}
where $m_{4}\gamma^{0}\gamma^{4}$ term destroys both the time-reversal
and spatial inversion symmetry but preserves their combination. The
term acts as the spin density wave order and can be realized by a
staggered Zeeman field \citep{Li2010np,sekine2014jpsj,Wang-prb-2022}.
Unlike the winding number, the topological $\theta$ term is defined
without assuming the sublattice symmetry and can be used for a system
with symmetry-breaking term $m_{4}\gamma^{0}\gamma^{4}$. In general,
the topological $\theta$ term can be obtained by integrating the
Chern-Simons three form over the Brillouin zone, $\frac{\theta}{2\pi}=\frac{1}{8\pi^{2}}\int_{BZ}d^{3}k\epsilon_{\alpha\beta\lambda}\mathrm{Tr}[A_{\alpha}\partial_{k_{\beta}}A_{\lambda}-\frac{2i}{3}A_{\alpha}A_{\beta}A_{\lambda}]$
\citep{Qi2008QFT,Essin2009OMP}, where $A_{\ell}^{\mu\nu}=i\langle u_{\mu}|\partial_{k_{\ell}}u_{\nu}\rangle$
is the non-abelian Berry connection defined from the Bloch function
of occupied band $|u_{\mu}\rangle$ and $|u_{\nu}\rangle$, the indices
$\alpha$, $\beta$, $\lambda$ run over 1,2,3. By choosing a proper
gauge to make the Berry connection being well-defined in the whole
momentum space, the topological $\theta$ term of $H_{\mathrm{surf}}^{4}$
can be evaluated as 
\begin{align}
\frac{\theta}{2\pi}= & \frac{\mathrm{sgn}(m_{4})}{2}\int d^{3}\mathbf{k}\mathcal{S}(k)\varpi_{3D},\label{eq:theta_term}
\end{align}
where $\mathcal{S}(k)=\frac{1}{2}\left(2-3\sin\varphi_{k}+\sin^{3}\varphi_{k}\right)$,
$\sin\varphi_{k}=\frac{1}{\sqrt{1+\left(h/m_{4}\right)^{2}}}$, the
integral is performed in the whole momentum space. For $k^{2}<k_{c}^{2}$,
although $\mathcal{S}(k)\ne0$, the winding number density vanishes
in this regime as $m(k)=0$; hence, we can restrict the integration
domain into $k^{2}>k_{c}^{2}$ in Eq. (\ref{eq:theta_term}). When
$\left|v\hbar k_{c}/m_{4}\right|\gg1$ or $|h/m_{4}|\gg1$, $\sin\varphi_{k}\to0$
and $\mathcal{S}(k)\to1$ for $k^{2}>k_{c}^{2}$, thus, we have
\begin{equation}
\frac{\theta}{2\pi}=\frac{1}{2}\mathrm{sgn}(m_{4})\omega_{3D},\label{eq:theta_term_1}
\end{equation}
which is quarter-quantized for a single gapless Dirac fermion. Compared
to the topological insulators, where half-quantized $\frac{\theta}{2\pi}$
corresponds to well-defined surface states, the quarter quantized
$\frac{\theta}{2\pi}$ is a pure bulk property of the single gapless
Dirac fermion as there is no localized boundary states. In addtion,
for the Wilson fermion or modified Dirac fermions in our previous
work, a finite small $m_{4}$ will open a gap in the energy spectrum,
and $\frac{\theta}{2\pi}$ is not quarter quantized any more \citep{Wang-prb-2022}.
However, for the Dirac fermion described by Eq. (\ref{eq:surface_4D-1}),
$\frac{\theta}{2\pi}$ is quarter-quantized even for a finite symmetry
breaking term $m_{4}\gamma^{0}\gamma^{4}$ once $\left|v\hbar k_{c}/m_{4}\right|\gg1$.
Moreover, it is noted that the topological $\theta$ term in Eq. (\ref{eq:theta_term})
or Eq. (\ref{eq:theta_term_1}) depends on the sign of $m_{4}$ even
in the limit of $m_{4}\to0$. Thus the $\theta$ field emerges in
the gapless Dirac system as a consequence of spontaneous symmetry
breaking induced by an extremely small field $m_{4}$. In transport
measurement, $\theta$ is tied to the slab Hall conductance and may
leads to some measurable effect in the experiments, such as the topological
Kerr and Faraday rotations \citep{maciejko2010prl,Tse2011prb,fu2021prr},
which provides an effective way to measure the magnetoelectric polarization
in solids. Recently, Chao et al \citep{Lei-2024-arxiv} proposed a
new strategy to detect the topological magneto-electrical effect by
capacitive detection of polarization response to magnetic field, which
might provide a higher accuracy than the Kerr and Faraday probes to
confirm the quantization of $\theta$ term in topological materials.

\section{Summary}

In summary, the $d$ dimensional single Dirac fermion can be realized
on the boundary of $d+1$ dimensional Chern insulators. For boundary
states, the chiral symmetry is locally preserved in the low energy
regime but violated at the high energy part. Based on the effective
continuum and lattice model, we studied the chiral anomaly effect
in the presence of electromagnetic field for these special states.
When the chemical potential resides in the chiral symmetric regime,
we recover the continuity equation of chiral current in the field
theory even for a finite chemical potential for both one and three
dimensional case. If we further put the chemical potential in the
high energy regime, where the chiral symmetry is not preserved any
more, the chiral current tends to be conserved as the two chiral fermions
are always mixed together near the fermi energy. Here we have established
a straightforward relation between the chiral symmetry and chiral
anomaly effect, which can help us have a deeper understanding on the
abstract quantum anomaly effect in condensed matter system. It might
also provide a new paradigm for the chiral gauge theory in particle
physics. At the last, we studied the magneto-electric effect associated
with the three-dimensional gapless Dirac fermion, which is quarter-quantized
in the presence of symmetry breaking term $m_{4}\gamma^{0}\gamma^{4}$
once $\left|v\hbar k_{c}/m_{4}\right|\gg1$.
\begin{acknowledgments}
This work was supported by the National Natural Science Foundation
of China under Grants No. 12304192 and No. 12504049; the Research Grants Council, University
Grants Committee, Hong Kong under Grant No. C7012-21G and No. 17301823;
the Quantum Science Center of Guangdong-Hong Kong-Macao Greater Bay
Area GDZX2301005; the Guangdong Province Introduced Innovative R\&D
Team Program under Grant No. 2023QN10X136; the Guangdong Basic and
Applied Basic Research Foundation under Grant No. 2024A1515010430
and No. 2023A1515140008; the Sichuan Science and Technology Program
under Grant No. 2024NSFSC1376; and the China Postdoctoral Science
Foundation under Grant No. 2023M740525.
\end{acknowledgments}

\appendix

\section{Solution of $H_{\mathrm{1d}}$\label{sec:Solution-of-E}}

For a nontrivial solution of $\Phi,$ the secular equation of $\xi$
and $E$ gives four solutions for $\xi(E)$ as 
\begin{equation}
\xi_{s\alpha}(E)=s\xi_{\alpha}=s\sqrt{\frac{(-1)^{\alpha-1}F-D}{2b^{2}}},\label{eq:solution_xi}
\end{equation}
where $s=\pm,\alpha=1,2,$ the parameters $D,F$ are defined as $D=v^{2}\hbar^{2}-2bm_{0}(k_{x},0),$
$F=\sqrt{D^{2}+4b^{2}\left(E^{2}-\left(m_{0}(k_{x},0)\right)^{2}\right)}.$

Considering the boundary condition $\Psi(k_{x},y=\pm\frac{L}{2})=0$
and parity symmetry of $H_{1d}$, we can further construct the ansatz
of $\Psi$ as \citep{Bai-arxiv-2024,zhou-prl-2008,Fu-nc-2024} 
\begin{equation}
\Psi(k_{x},y)=\begin{pmatrix}\tilde{c}_{+}g_{+}(y)+\tilde{c}_{-}g_{-}(y)\\
\tilde{d}_{+}g_{+}(y)+\tilde{d}_{-}g_{-}(y)
\end{pmatrix},\label{eq:ansatz_wave_function}
\end{equation}
where $g_{+}=\frac{\cos\xi_{1}y}{\cos\left(\xi_{1}L/2\right)}-\frac{\cos\xi_{2}y}{\cos\left(\xi_{2}L/2\right)}$,
$g_{-}=\frac{\sin\xi_{1}y}{\sin\left(\xi_{1}L/2\right)}-\frac{\sin\xi_{2}y}{\sin\left(\xi_{2}L/2\right)}$,
the subscripts $\pm$ represent even ($+$) and odd ($-$) parity,
respectively. Then, the boundary condition $\Psi(k_{x},y=\pm\frac{L}{2})=0$
is naturally satisfied in this basis as $g_{\pm}(y=\pm\frac{L}{2})=0$.
To find the relation between $\xi$ and $E$, plugging the ansatz
in Eq. (\ref{eq:ansatz_wave_function}) into the eigen equation $H_{1d}\Psi=E\Psi$,
after some algebra, we can obtain a system of linear equations involving
$\tilde{c}_{\pm}$ and $\tilde{d}_{\pm}$. The nontrivial solution
of $\tilde{c}_{\pm}$ and $\tilde{d}_{\pm}$ leads to a secular equation,
\begin{align}
\frac{\tan\frac{\xi_{1}L}{2}}{\tan\frac{\xi_{2}L}{2}}+\frac{\tan\frac{\xi_{2}L}{2}}{\tan\frac{\xi_{1}L}{2}}= & \frac{\xi_{2}\mathcal{E}(k_{x},\xi_{1})}{\xi_{1}\mathcal{E}(k_{x},\xi_{2})}+\frac{\xi_{1}\mathcal{E}(k_{x},\xi_{2})}{\xi_{2}\mathcal{E}(k_{x},\xi_{1})},\label{eq:secular equation}
\end{align}
where $\mathcal{E}(k_{x},\xi)=E-m_{0}(k_{x},\xi)$. Eq. (\ref{eq:secular equation})
gives energy dispersions for states labeled as $E_{+}$ and $E_{-}$,
respectively,
\begin{align}
E_{+} & =m_{0}-bk_{x}^{2}+b\xi_{1}\xi_{2}\frac{\xi_{2}\tan\frac{\xi_{2}L}{2}-\xi_{1}\tan\frac{\xi_{1}L}{2}}{\xi_{2}\tan\frac{\xi_{1}L}{2}-\xi_{1}\tan\frac{\xi_{2}L}{2}},\label{eq:Ep}\\
E_{-} & =-E_{+}.\label{eq:Em}
\end{align}
As $\{H_{1d},\sigma_{x}\}=0$, the eigen states of $E_{\pm}$ are
tied to each other as $\Psi_{-}=\sigma_{x}\Psi_{+}$; hence, we can
firstly focus on state of $E_{+}$ in the following discussion. After
a straightforward derivation, the eigen state of $E_{+}$ can be found
as
\begin{align}
\Psi_{+}(k_{x},y)= & \tilde{n}\begin{pmatrix}v\hbar g_{+}\\
-b\eta g_{-}
\end{pmatrix},\label{eq:wave_func_1}
\end{align}
where $\eta=\frac{\xi_{1}^{2}-\xi_{2}^{2}}{\xi_{1}\cot\left(\xi_{1}L/2\right)-\xi_{2}\cot\left(\xi_{2}L/2\right)},$
$\tilde{n}$ is the normalization constant.

To find the solution of $E_{+}$ , we consider several cases for $\xi_{1,2}$.
If $\mathrm{Im}\xi_{1}>0>\mathrm{Im}\xi_{2}$, $\lim_{L\to+\infty}\tan\frac{\xi_{1}L}{2}=i$,
$\lim_{L\to+\infty}\tan\frac{\xi_{2}L}{2}=-i$, then $\lim_{L\to+\infty}E_{+}=m_{0}(k_{x},0)-b\xi_{1}\xi_{2}.$
Together with the solution of $\xi_{1,2}$ in Eq. (\ref{eq:solution_xi}),
we have $E_{+}=0$, and

\begin{align}
\xi_{1,2} & =\begin{cases}
\xi_{0}\left(\pm i+\sqrt{u_{m}-1}\right), & u_{m}>1\\
\xi_{0}\left(\pm i+i\sqrt{1-u_{m}}\right), & 0<u_{m}<1.
\end{cases}
\end{align}
with $u_{m}=4b\left(m_{0}-bk_{x}^{2}\right)/v^{2}\hbar^{2}$ and $\xi_{0}=\frac{v\hbar}{2|b|}$
dimensionless parameters.

Similarly, for $\mathrm{Im}\xi_{1,2}>0$, $\lim_{L\to+\infty}\tan\xi_{1,2}L/2=i$,
we have $E_{+}=0$, and 
\begin{align}
\xi_{1,2} & =\begin{cases}
\xi_{0}\left(i\pm\sqrt{u_{m}-1}\right), & u_{m}>1\\
\xi_{0}\left(i\pm i\sqrt{1-u_{m}}\right), & 0<u_{m}<1
\end{cases}
\end{align}
In these two cases, the imaginary part of $\xi_{1,2}$ are nonzero,
the edge states are localized near the boundary of the ribbon. If
$\mathrm{Im}\xi_{1}=0$, or $\mathrm{Im}\xi_{2}=0$, we will have
$E_{+}=m_{0}(k_{x},0)$ and $u_{m}<0$.

In summary, we obtain the two solutions of $E_{\pm}$ as
\begin{align}
E_{+} & =\begin{cases}
0, & m_{0}(k_{x},0)b>0\\
m_{0}(k_{x},0) & m_{0}(k_{x},0)b<0
\end{cases}\label{eq:E+}\\
E_{-} & =-E_{+}.\label{eq:E-}
\end{align}
For the complete solution for $E_{\pm}$, we can solve Eq. (\ref{eq:solution_xi})
and Eq. (\ref{eq:secular equation}) numerically. As shown in Fig.
\ref{fig:Epm}, there is a good agreement between the numerical results
(filled and open circles) and analytical results (blue lines) in Eq.
(\ref{eq:E+}) and Eq. (\ref{eq:E-}) for the lowest two states near
the half-filling. Moreover, for the remaining states, the eigenergies
$E_{\pm}$ can be quantitatively described by $E_{n,\pm}=\pm\sqrt{v^{2}\hbar^{2}k_{y,n}^{2}+\left(m(k_{x},k_{y,n})\right)^{2}}$,
where $k_{y,n}=\frac{n\pi}{L}$ is the discrete wave vectors and $n$
is positive integer.

\setcounter{figure}{0}
\renewcommand{\thefigure}{A\arabic{figure}} 

\begin{figure}
\begin{centering}
\includegraphics[width=7cm]{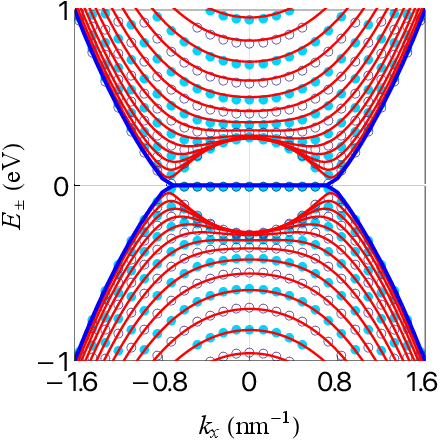}
\par\end{centering}
\caption{\label{fig:Epm}Energy spectrum of the $E_{\pm}$ in the continuum
model given by Eq. (\ref{eq:secular equation}), where the filled
and open circles represent the eigenvalues of $E_{+}$ and $E_{-}$,
respectively. The blue lines denote the analytical results in Eq.
(\ref{eq:E+}) and Eq. (\ref{eq:E-}). The red lines are given by
$E_{n,\pm}=\pm\sqrt{v^{2}\hbar^{2}k_{y,n}^{2}+\left(m(k_{x},k_{y,n})\right)^{2}}$.
The length of the ribbon is chosen as $L=31\,\mathrm{nm}$ along $y$
direction. The model parameters are set as $v\hbar=0.44\,\mathrm{eV}\cdot\mathrm{nm}$,
$m_{0}=0.28\,\mathrm{eV}$, $b=0.5\,\mathrm{eV}\cdot\mathrm{nm}^{2}$.}
\end{figure}

\section{Lattice Model\label{sec:Lattice-Model}}

Now, let us turn to a more realistic model. In this part, we put the
CIs on a lattice as 
\begin{align}
H_{L} & =\lambda_{L}\left(\sin k_{x}a\sigma_{x}+\sin k_{x}a\sigma_{y}\right)+m_{0}(\mathbf{k})\sigma_{z},\label{eq:hamiltonian_lattice}
\end{align}
where $m_{0}(\mathbf{k})=m_{0}-4t_{L}(\sin^{2}\frac{k_{x}a}{2}+\sin^{2}\frac{k_{y}a}{2})$
is the momentum dependent Dirac mass term, $t_{L},\lambda_{L},m_{0}$
are model parameters, $a$ is the lattice spacing. In the continuum
limit, $a\to0$, $\sin k_{i}a\sim k_{i}a,$$\cos k_{i}a\sim1-\frac{k_{i}^{2}a^{2}}{2}$,
one can restore the effective model in Eq. (\ref{eq:continuum model}),
where $t_{L}=\frac{b}{a^{2}},\lambda_{L}=\frac{v\hbar}{a}$.

To find the edge Hamiltonian, we consider open boundary condition
along $y$ direction, then
\begin{align}
H_{L}= & \sum_{k_{x}j_{y}}c_{k_{x}j_{y}}^{\dagger}\left(M_{0}(k_{x})\sigma_{z}+\lambda_{L}\sin k_{x}a\sigma_{x}\right)c_{k_{x}j_{y}}\\
 & +\sum_{k_{x},j_{y}}\left(c_{k_{x}j-1}^{\dagger}Tc_{k_{x}j}+c_{k_{x}j+1}^{\dagger}T^{\dagger}c_{k_{x}j}\right),\nonumber 
\end{align}
where $M_{0}(k_{x})=m-2t_{L}\left(1+2\sin^{2}\frac{k_{x}a}{2}\right)$
and $T=-\frac{\lambda_{L}}{2}i\sigma_{y}+t_{L}\sigma_{z}$. We can
further divide $H_{L}$ into two parts as
\begin{equation}
H_{L}=H_{1d}+H_{x},
\end{equation}
where
\begin{align}
H_{1d} & =\sum_{k_{x}j_{y}}\left(c_{k_{x}j_{y}}^{\dagger}M_{0}(k_{x})c_{k_{x}j_{y}}+c_{k_{x}j-1}^{\dagger}Tc_{k_{x}j}+h.c.\right),\\
H_{x} & =\sum_{k_{x},j_{y}}c_{k_{x}j_{y}}^{\dagger}\lambda_{L}\sin k_{x}a\sigma_{x}c_{k_{x}j_{y}}.
\end{align}
We first focus on $H_{1d}$ , which satisfies the following eigen
equation
\begin{equation}
H_{1d}\phi=E\phi,\label{eq:eigen-eq-lattice}
\end{equation}
or
\begin{equation}
T^{\dagger}\phi_{\ell_{y}-1}+\left(M_{0}(k_{x})\sigma_{z}-E\right)\phi_{\ell_{y}}+T\phi_{\ell_{y}+1}=0\label{eq:eigen-eq-lattice-2}
\end{equation}
with $\phi(\mathbf{k})=\oplus_{\ell_{y}}\phi_{\ell_{y}}(\mathbf{k})$.

Setting a trial solution $\phi_{\ell_{y}}=e^{i\xi\ell_{y}}\phi$,
one has 
\begin{equation}
\left[T^{\dagger}e^{-i\xi}+(M_{0}(k_{x})\sigma_{z}-E)+Te^{i\xi}\right]\phi_{\ell_{y}}=0.
\end{equation}
For a nontrivial solution of $\phi,$ $\xi$ and $E$ satisfy the
secular equation
\begin{equation}
\mathrm{det}\left[T^{\dagger}e^{-i\xi}+(M_{0}(k_{x})\sigma_{z}-E)+Te^{i\xi}\right]=0,
\end{equation}
which results in
\begin{widetext}
\begin{align}
\cos\xi_{\alpha} & =\frac{-2M_{0}(k_{x})t_{L}+(-1)^{\alpha-1}\sqrt{\left(M_{0}(k_{x})\lambda_{L}\right)^{2}-(4t_{L}^{2}-\lambda_{L}^{2})(\lambda_{L}^{2}-E^{2})}}{4t_{L}^{2}-\lambda_{L}^{2}},\alpha=1,2,\\
\sin\xi_{\alpha}^{s} & =s\sqrt{1-\cos^{2}\xi_{\alpha}},s=\pm.\nonumber 
\end{align}
\end{widetext}

According to the boundary condition $\Phi(\ell_{z}=\pm L/2)=0$, we
can reconstruct the eigen states as 
\begin{equation}
\phi_{\ell_{y}}^{s}=\begin{pmatrix}\tilde{c}_{+}g_{+}(\ell_{y})+\tilde{c}_{-}g_{-}(\ell_{y})\\
\tilde{d}_{+}g_{+}(\ell_{y})+\tilde{d}_{-}g_{-}(\ell_{y})
\end{pmatrix},
\end{equation}
where $g_{+}(\ell_{y})=\frac{\cos\xi_{1}\ell_{y}}{\cos\xi_{1}L/2}-\frac{\cos\xi_{2}\ell_{y}}{\cos\xi_{2}L/2}$,
$g_{-}(\ell_{y})=\frac{\sin\xi_{1}\ell_{y}}{\sin\xi_{1}L/2}-\frac{\sin\xi_{2}\ell_{y}}{\sin\xi_{2}L/2}$.
Plugging $\phi_{\ell_{y}}^{s}$ into the eigen equation in Eq. (\ref{eq:eigen-eq-lattice-2}),
and after some algebra, one have
\begin{align}
\frac{\tan\xi_{1}L/2}{\tan\xi_{2}L/2}+\frac{\tan\xi_{2}L/2}{\tan\xi_{1}L/2}= & \frac{\sin\xi_{2}\mathcal{E}_{+}(k_{x},\xi_{1})}{\sin\xi_{1}\mathcal{E}_{+}(k_{x},\xi_{2})}+\frac{\sin\xi_{1}\mathcal{E}_{+}(k_{x},\xi_{2})}{\sin\xi_{2}\mathcal{E}_{+}(k_{x},\xi_{1})},
\end{align}
with $\mathcal{E}_{+}(k_{x},\xi_{\alpha})=E-M(k_{x})-2t_{L}\cos\xi_{\alpha}$.
This secular equation gives energy dispersions for electrons designated
as $E_{+}$ and $E_{-}$, respectively,
\begin{align}
E_{+} & =M_{0}(k_{x})+2t_{L}\frac{\cos\xi_{1}\sin\xi_{2}-x\cos\xi_{2}\sin\xi_{1}}{\sin\xi_{2}-x\sin\xi_{1}},\\
E_{-} & =-E_{+},
\end{align}
with $x=\tan\frac{\xi_{2}L}{2}/\tan\frac{\xi_{1}L}{2}.$ The expressions
for the corresponding wave functions are given by
\begin{align}
\Phi_{+}(k_{x},y)= & \tilde{c}\begin{pmatrix}\lambda_{L}g_{+}\\
-t_{L}\eta g_{-}
\end{pmatrix},\\
\Phi_{-}(k_{x},y)= & \tilde{c}\begin{pmatrix}-t_{L}\eta g_{-}\\
\lambda_{L}g_{+}
\end{pmatrix}.
\end{align}
The projection on lowest two states gives 'edge' Hamiltonian $(H_{e})_{ij}=\sum_{\ell_{y}}\Phi_{i}^{\dagger}(\ell_{y})H(k_{x},\ell_{y})\Phi_{j}(\ell_{y})$,
which results in
\begin{align}
H_{e}= & m(k_{x})\sigma_{z}+\lambda_{L}\sin k_{x}a\sigma_{x}
\end{align}
where 
\begin{equation}
m(k_{x})=\begin{cases}
0, & 0<M_{0}(k_{x})<4t_{L},\\
M_{0}(k_{x})+2t_{L}, & \mathrm{otherwise}.
\end{cases}
\end{equation}

Then, the eigen energies of $H_{e}$ can be found as 
\begin{equation}
E_{\pm}=\pm\sqrt{m(k_{x})^{2}+\lambda_{L}^{2}\sin^{2}k_{x}a}.\label{eq:Eigen_energy_L}
\end{equation}
The eigen states with small wave vector localized along two different
edges as depicted in Fig. \ref{fig:(a-b)-Spatial-distributions-S1}(a),
and and can be well described by Eq. (\ref{eq:WF_Edge States}). For
a large wave vector, as shown in Fig. \ref{fig:(a-b)-Spatial-distributions-S1}(b),
the edge states evolved into bulk states, and the corresponding density
$|\phi_{s}^{\pm}|^{2}$ distributed in the whole bulk region of the
ribbon.

\setcounter{figure}{0}
\renewcommand{\thefigure}{B\arabic{figure}} 

\begin{figure}
\begin{centering}
\includegraphics[width=7.5cm]{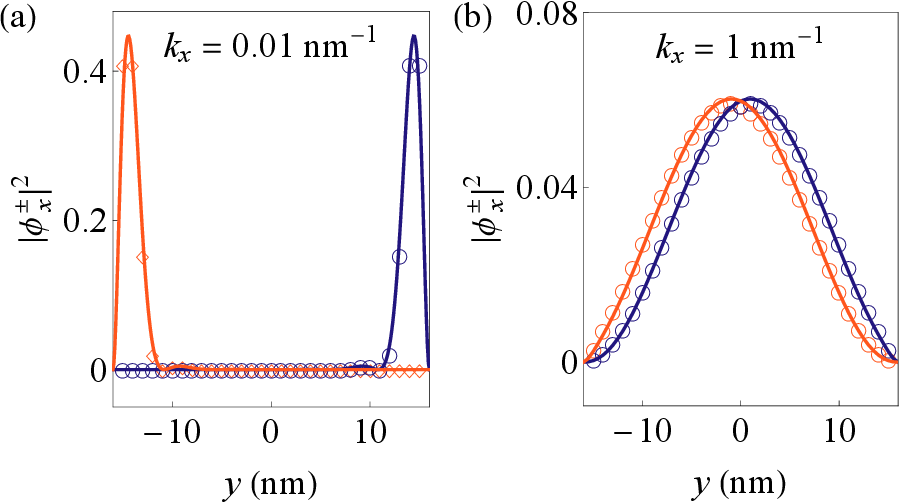}
\par\end{centering}
\caption{\label{fig:(a-b)-Spatial-distributions-S1}Density distributions ($|\phi_{x}^{\pm}(k_{x},y)|^{2}$)
of the two energy bands near the half-filling along $y$ direction
for (a) $k_{x}=0.01\,\mathrm{nm^{-1}}$ and (b) $k_{x}=1\,\mathrm{nm}^{-1}$.
The circles and solid lines are the results of the lattice model and
continuum model, respectively. The length of the ribbon is chosen
as $L=31\,\mathrm{nm}$. The model parameters are set as $v\hbar=0.44\,\mathrm{eV}\cdot\mathrm{nm}$,
$m_{0}=0.28\,\mathrm{eV}$, $b=0.5\,\mathrm{eV}\cdot\mathrm{nm}^{2}$,
the lattice spacing is assumed as $1\,\mathrm{nm}$.}
\end{figure}

\section{\label{sec:Zero-Modes}Zero Modes}

In last section, we consider a general case for $H_{1d}$, where the
eigen energy $E$ satisfies two transcendental equations. Here we
consider a special case with $E=0$ and semi-infinite length along
$y$ direction. The corresponding eigen equation becomes 
\begin{equation}
\left\{ \left(2t_{L}\cos\xi+M_{0}(k_{x})\right)\sigma_{z}+\lambda_{L}\sin\xi\sigma_{y}\right\} \phi_{\ell_{y}}=0,\label{eq:eigen-eq-zero}
\end{equation}
multiplied by $\sigma_{z}$ in both sides of Eq. (\ref{eq:eigen-eq-zero}),
one has
\begin{equation}
\left\{ 2t_{L}\cos\xi+M_{0}(k_{x})-i\lambda_{L}\sin\xi\sigma_{x}\right\} \phi_{\ell_{y}}=0.\label{eq:eigen-eq-zero-2}
\end{equation}
In terms of eigenstates of $\sigma_{x}$, $\sigma_{x}\zeta_{s_{x}}=s_{x}\zeta_{s_{x}}$
with $s_{x}=\pm1$, $\phi_{\ell_{y}}=\varphi_{\ell_{y}}\zeta_{s_{x}}$,
Eq. (\ref{eq:eigen-eq-zero-2}) becomes a scaler equation 
\begin{equation}
2t_{L}\cos\xi+M_{0}(k_{x})-is_{x}\lambda_{L}\sin\xi=0,
\end{equation}
which gives two solutions of $\xi$ as 
\begin{equation}
e^{i\xi_{\pm}}=\frac{-M_{0}(k_{x})\pm\sqrt{M_{0}^{2}(k_{x})-\left(4t_{L}^{2}-\lambda_{L}^{2}\right)}}{2t_{L}-s_{x}\lambda_{L}}.
\end{equation}

Then, the general solution for the eigen states can be expressed as
\begin{equation}
\Psi(\ell_{y},k_{x})=\sum_{s_{x}}\left(c_{s_{x},+}e^{i\xi_{+}\ell_{y}}+c_{s_{x},-}e^{i\xi_{-}\ell_{y}}\right)\varphi(k_{x})\zeta_{s_{x}}.
\end{equation}

Considering the boundary condition $\Psi(\ell_{y}=0,k_{x})=0$ and
$\Psi(\ell_{y}=+\infty,k_{x})=0$, one has $c_{s_{x},+}=-c_{s_{x},-}$,
and $|e^{i\xi_{\pm}}|<1$. Hence,
\begin{equation}
\left|e^{i(\xi_{+}+\xi_{-})}\right|=\left|\frac{4t_{L}^{2}-\lambda_{L}^{2}}{(2t_{L}-s_{x}\lambda_{L})^{2}}\right|<1,
\end{equation}
which gives $s_{x}=-\mathrm{sgn}(t_{L}\lambda_{L}).$ Without loss
generality, we assume $\mathrm{sgn}(t_{L}\lambda_{L})>0$, then, $s_{x}=-1$
and 
\begin{equation}
\left|M_{0}(k_{x})\mp\sqrt{M_{0}^{2}(k_{x})-\left(4t_{L}^{2}-\lambda_{L}^{2}\right)}\right|<2t_{L}+\lambda_{L},
\end{equation}
which leads to 
\begin{equation}
0<m_{0}-4t_{L}\sin^{2}\frac{k_{x}a}{2}<4t_{L}.
\end{equation}

Therefore, we have demonstrated that the zero mode of CIs can only
exist in a finite momentum regime ($0<m_{0}-4t_{L}\sin^{2}\frac{k_{x}a}{2}<4t_{L}$),
in this regime, the chiral symmetry is locally preserved for the edge
states. However, when moving out this regime, there is no solution
for zero mode. The edge states evolve into the bulk states, and the
chiral symmetry is explicitly broken in the high energy regime. Such
a property exactly satisfies the criteria of quantum anomalous semimetal
in one dimensions. In the main text, we further studied the quantum
anomaly for this edge state to confirm its peculiar property as a
quantum anomalous semimetal.

\section{\label{sec:Linear-response-theory}Linear response theory of Pseudo-scalar
condensation}

In this section, we derive the expectation value of Pseudo-scalar
condensation from the linear response theory. In the linear response
theory, the expectation value of $\bar{\psi}\hat{\mathcal{O}}\psi$
can be computed as \citet{Wang-prb-2022}

\begin{align*}
\langle\bar{\psi}\hat{\mathcal{O}}\psi\rangle= & E_{x}\lim_{\Omega\to0}\left(\Lambda^{RA}+\Lambda^{RR}-\Lambda^{AA}\right)
\end{align*}
where
\begin{align*}
\Lambda^{RA} & =\lambda_{L}\sum_{k_{x}}\int_{-\infty}^{\infty}d\omega\frac{n_{F}(\omega^{\prime})-n_{F}(\omega)}{\Omega}\mathrm{Tr}\left[\hat{\mathcal{O}}G_{k_{x}}^{R}(\omega^{\prime})\hat{j}_{x}G_{k_{x}}^{A}(\omega)\right],\\
\Lambda^{RR} & =\frac{\lambda_{L}}{\Omega}\sum_{k_{x}}\int_{-\infty}^{\infty}d\omega n_{F}(\omega)\mathrm{Tr}\left[\hat{\mathcal{O}}G_{k_{x}}^{R}(\omega^{\prime})\hat{j}_{x}G_{k_{x}}^{R}(\omega)\right],\\
\Lambda^{AA} & =\frac{\lambda_{L}}{\Omega}\sum_{k_{x}}\int_{-\infty}^{\infty}d\omega n_{F}(\omega^{\prime})\mathrm{Tr}\left[\hat{\mathcal{O}}G_{k_{x}}^{A}(\omega^{\prime})\hat{j}_{x}G_{k_{x}}^{A}(\omega)\right],
\end{align*}
where $\lambda_{L}=\frac{e\hbar}{2\pi L_{x}}$, $\omega^{\prime}=\omega+\Omega$,
$\hat{j}_{x}=\frac{1}{\hbar}\frac{\partial\mathds{h}}{\partial k_{x}}$
is the velocity operator, and $G_{k_{x}}^{R/A}(\omega)=i\left[\left(\omega\pm i\delta\right)\gamma^{0}-\mathds{h}\right]^{-1}$
are the retarded/advanced Green\textquoteright s functions. $n_{F}(\omega)=(\mu-\omega)$
is the Fermi-Dirac distribution function at zero temperature. $L_{x}$
is the length of the effective one-dimensional system. For $\mathds{h}=\gamma^{1}h_{1}(k_{x})+m(k_{x})$
and $\mathcal{\hat{O}}=2im(k_{x})\gamma^{5}/\hbar$, after a tedious
but straightforward calculation, we obtain the response functions
as
\begin{align*}
\lim_{\Omega\to0}\Lambda^{RA}= & -\frac{e}{\pi\hbar}\frac{m(k_{x})\left(m(k_{x})-m^{\prime}(k_{x})h_{1}(k_{x})\right)}{\mu^{2}|\varepsilon^{\prime}(k_{x})|}\bigg|_{k_{x}=k_{F}},
\end{align*}
\[
\lim_{\Omega\to0}\left(\Lambda^{RR}-\Lambda^{AA}\right)=-\lim_{\Omega\to0}\Lambda^{RA}-\frac{eE_{x}}{\pi\hbar}\frac{h_{1}(k_{x})}{\varepsilon(k_{x})}\bigg|_{k_{x}=k_{F}}^{k_{\partial}}
\]
with $\varepsilon(k_{x})=\sqrt{\left(m(k_{x})\right)^{2}+v^{2}\hbar^{2}k_{x}^{2}}$.

Then, the expectation value of Pseudo-scalar condensation at the dc
limit becomes
\begin{align*}
\langle\bar{\psi}\frac{2im(k_{x})}{\hbar}\gamma^{5}\psi\rangle= & -\frac{eE_{x}}{\pi\hbar}\frac{h_{1}(k_{x})}{\varepsilon(k_{x})}\bigg|_{k_{x}=k_{F}}^{k_{\partial}},
\end{align*}
which is consistent with results from the perturbed eigenstates in
a constant electric field.


\begin{thebibliography}{Stephanov (2012)}
\bibitem[adler(1969)]{adler1969} S. L. Adler, Axial-vector vertex
in spinor electrodynamics, Phys. Rev. \textbf{177}, 2426 (1969).

\bibitem[bell(1969)]{Bell1969} J. S. Bell and R. Jackiw, A PCAC puzzle:
$\pi^{0}\to\gamma\gamma$ in the $\sigma$-model, II Nuovo Cimento
A \textbf{60}, 47 (1969).

\bibitem[fujikawa(1979)]{fujikawa1979} K. Fujikawa, Path-Integral
Measure for Gauge-Invariant Fermion Theories, Phys. Rev. Lett. 42,
1195 (1979).

\bibitem[peskin(1995)]{peskin_book}M. E. Peskin and D. V. Schroeder,
An Introduction to Quantum Field Theory (Perseus Books Publishing
LLC, Massachusetts, 1995).

\bibitem[Azee(2010)]{Azee-book}A. Zee, \emph{Quantum Field Theory
in a Nutshell} (Princeton University Press, Princeton, NJ, 2010).

\bibitem[Nielsen(1983)]{Nielsen_plb_1983}H. Nielsen and M. Ninomiya,
The Adler-Bell-Jackiw anomaly and Weyl fermions in a crystal, Phys.
Lett. B 130, 389 (1983).

\bibitem[Stephanov(2012)]{Stephanov-prl-2012}M. A. Stephanov and
Y. Yin, Chiral Kinetic Theory, Phys. Rev. Lett. \textbf{109}, 162001
(2012.)

\bibitem[armitage(2018)]{armitage2018rmp}N. P. Armitage, E. J. Mele,
and A. Vishwanath, Weyl and Dirac semimetals in three-dimensional
solids, Rev. Mod. Phys. \textbf{90}, 015001 (2018).

\bibitem[lu(2017)]{lu-FP-2017}H. Z. Lu and S.-Q. Shen, Quantum transport
in topological semimetals under magnetic fields, Front. Phys. 12,
127201 (2017).

\bibitem[DTSon(2013)]{DTSon_prb_2013}D. T. Son and B. Z. Spivak,
Chiral anomaly and classical negative magnetoresistance of Weyl metals,
Phys. Rev. B 88, 104412 (2013).

\bibitem[kim(2013)]{kim-prl-2013}H. J. Kim, K. S. Kim, J. F. Wang,
M. Sasaki, N. Satoh, A. Ohnishi, M. Kitaura, M. Yang, and L. Li, Dirac
Versus Weyl Fermions in Topological Insulators: Adler-Bell-Jackiw
Anomaly in Transport Phenomena, Phys. Rev. Lett. \textbf{111}, 246603
(2013).

\bibitem[xiong(2015)]{xiong-science-2015}J. Xiong, S. K. Kushwaha,
T. Liang, J. W. Krizan, M. Hirschberger, W. Wang, R. J. Cava, and
N. P. Ong, Evidence for the chiral anomaly in the Dirac semimetal
$\mathrm{Na_{3}Bi}$, Science \textbf{350}, 413 (2015).

\bibitem[zhang(2016)]{zhang-nc-2016}C. L. Zhang, S. Y. Xu, I. Belopolski,
Z. Yuan, Z. Lin, B. Tong, G. Bian, N. Alidoust, C. C. Lee, S. M. Huang,
et al., Signatures of the Adler-Bell-Jackiw chiral anomaly in a Weyl
fermion semimetal, Nat. Commun. \textbf{7}, 10735 (2016).

\bibitem[li\_nc(2016)]{Li-nc-2016}H. Li, H. He, H. Z. Lu, H. Zhang,
H. Liu, R. Ma, Z. Fan, S. Q. Shen, and J. Wang, Negative magnetoresistance
in Dirac semimetal $\mathrm{Cd_{3}As_{2}}$, Nat. Commun. \textbf{7},
10301 (2016).

\bibitem[liang(2018)]{Liang-prx-2018}S. Liang, J. Lin, S. Kushwaha,
J. Xing, N. Ni, R. J. Cava, and N. P. Ong, Experimental Tests of the
Chiral Anomaly Magnetoresistance in the Dirac-Weyl Semimetals $\mathrm{Na_{3}Bi}$
and GdPtBi, Phys. Rev. X \textbf{8}, 031002 (2018).

\bibitem[li\_np(2016)]{Li16natphys} Q. Li, D. E. Kharzeev, C. Zhang,
Y. Huang, I. Pletikosic, A. V. Fedorov, R. D. Zhong, J. A. Schneeloch,
G. D. Gu, and T. Valla, Chiral magnetic effect in $\mathrm{ZrTe}_{5}$,
Nat. Phys. \textbf{12}, 550 (2016).

\bibitem[burkov(2014)]{Burkov_prl_2014}A. Burkov, Chiral Anomaly
and Diffusive Magnetotransport in Weyl Metals, Phys. Rev. Lett. \textbf{113},
247203 (2014).

\bibitem[huang(2017)]{huang-prb-2017}Z. M. Huang, J. Zhou, and S.
Q. Shen, Topological responses from chiral anomaly in multi-Weyl semimetals,
Phys. Rev. B 96, 085201 (2017).

\bibitem[fujikawa(1980)]{Fujikawa-prd-1980}K. Fujikawa, Path integral
for gauge theories with fermions, Phys. Rev. D \textbf{21}, 2848 (1980).

\bibitem[Wilson(1977)]{Wilson1977book}K. G. Wilson, in New Phenomena
in Subnuclear Physics, edited by A. Zichichi, (Plenum, New York, 1977).

\bibitem[Nilsen(1981)]{Nielsen1981PLB}H. B. Nielsen, and M. Ninomiya,
No go theorem for regularizing chiral fermions, Phys. Lett. \textbf{B105}
219 (1981).

\bibitem[jackiw(1969)]{jackiw1969pr}R. Jackiw and K. Johnson, Anomalies
of the axial-vector current, Phys. Rev. \textbf{182}, 1459 (1969).

\bibitem[kaplan(1992)]{Kaplan-plb-1992}D. B. Kaplan, A method for
simulating chiral fermions on the lattice, Phys. Lett. B \textbf{288},
342-347(1992).

\bibitem[Rothe(1998)]{Rothe1998book}H.\LyXThinSpace J. Rothe, Lattice
Gauge Theories (World Scientific, Singapore, 1998).

\bibitem[Kaplan\_1(2024)]{Kaplan-prl-2024}D. B. Kaplan, Chiral Gauge
Theory at the Boundary between Topological Phases, Phys. Rev. Lett.
\textbf{132}, 141603 (2024).

\bibitem[Kaplan\_2(2024)]{Kaplan-prl-2024-1}D. B. Kaplan and S. Sen,
Weyl Fermions on a Finite Lattice, Phys. Rev. Lett. \textbf{132},
141604 (2024).

\bibitem[Neuberger(1998)]{Neuberger-prl-1998}H. Neuberger, A Practical
Implementation of the Overlap Dirac Operator, Phys. Rev. Lett. \textbf{81},
4060 (1998).

\bibitem[Callan(1995)]{Callan-npb-1985}C. G. Callan, J. A. Harvey,
Anomalies and fermion zero modes on strings and domain walls, Nucl.
Phys. B \textbf{250}, 427 (1985).

\bibitem[Qi(2008)]{Qi2008QFT}X. L. Qi, T. L. Hughes, and S. C. Zhang,
Topological field theory of time-reversal invariant insulators, Phys.
Rev. B \textbf{78}, 195424 (2008).

\bibitem[Qi(2011)]{Qi2011rmp}X. L. Qi, and S. C. Zhang, Topological
insulators and superconductors, Rev. Mod. Phys. \textbf{83}, 1057
(2011).

\bibitem[Moore(2010)]{Moore2010nature}J. E. Moore, The birth of topological
insulators, Nature \textbf{464}, 194 (2010)

\bibitem[Hasan(2010)]{Hasan2010rmp}M. Z. Hasan, and C. L. Kane, Colloquium:
topological insulators, Rev. Mod. Phys. \textbf{82}, 3045 (2010)

\bibitem[SQS(2010)]{SQS}S. Q. Shen, \textit{Topological Insulators:
Dirac Equation in Condensed Matter}, 2nd ed. (Springer, Singapore,
2017).

\bibitem[Fu(2022)]{Fu-npj-2022}B. Fu, J. Y. Zou, Z. A. Hu, H. W.
Wang, and S. Q. Shen, Quantum Anomalous Semimetals, npj Quantum Mater.
\textbf{7}, 94 (2022).

\bibitem[Mogi(2022)]{Mogi-np-2022}M. Mogi, Y. Okamura, M. Kawamura,
R. Yoshimi, K. Yasuda, A. Tsukazaki, K. S. Takahashi, T. Morimoto,
N. Nagaosa, M. Kawasaki, Y. Takahashi, and Y. Tokura, Experimental
signature of parity anomaly in semi-magnetic topological insulator,
Nat. Phys. \textbf{18}, 390 (2022).

\bibitem[Zou(2022)]{Zou-prb-2022}J. Y. Zou, B. Fu, H. W. Wang, Z.
A. Hu, S. Q. Shen, Half-quantized Hall effect and power law decay
of edge-current distribution, Phys. Rev. B \textbf{105}, L201106 (2022).

\bibitem[Zou(2022)]{zou-prb-2023}J. Y. Zou, R. Chen, B. Fu, H. W.
Wang, Z. A. Hu, and S. Q. Shen, Half-quantized Hall effect at the
parity-invariant Fermi surface, Phys. Rev. B \textbf{107}, 125153
(2023).

\bibitem[SQS(2024)]{SQS-HQH}S.-Q. Shen, Half quantized Hall effect,
Coshare Science \textbf{2}, 1 (2024).

\bibitem[Wang(2024)]{Wang-prb-2024}H.-W. Wang, B. Fu, and S.-Q. Shen,
Signature of parity anomaly: Crossover from one half to integer quantized
Hall conductance in a finite magnetic field. Phys. Rev. B \textbf{109},
075113 (2024).

\bibitem[Chen(2024)]{Chen-2024-scpma}R. Chen, S.-Q. Shen, On the
half-quantized Hall conductance of massive surface electrons in magnetic
topological insulator films. Sci. China Phys. Mech. Astron. \textbf{67},
267011 (2024).

\bibitem[Fu(2025)]{Fu-arxiv}B. Fu and S.-Q. Shen, $\mathbb{Z}/2$
topological invariants and the half quantized Hall effect. Commun
Phys \textbf{8}, 2 (2025).

\bibitem[Bai(2024)]{Bai-arxiv-2024}K.-Z. Bai, B. Fu, and S.-Q. Shen,
Dirac fermions and topological phases in magnetic topological insulator
films, SciPost Phys. \textbf{17}, 146 (2024).

\bibitem[Wang(2022)]{Wang-prb-2022}H.-W. Wang, B. Fu, J.-Y. Zou,
Z.-A. Hu, and S.-Q. Shen, Fractional electromagnetic response in a
three-dimensional chiral anomalous semimetal, Phys. Rev. B \textbf{106},
045111 (2022).

\bibitem[Zhou(2008)]{zhou-prl-2008}B. Zhou, H.-Z. Lu, R.-L. Chu,
S.-Q. Shen, and Q. Niu, Finite Size Effects on Helical Edge States
in a Quantum Spin-Hall System, Phys. Rev. Lett. \textbf{101}, 246807
(2009).

\bibitem[Fu(2024)]{Fu-nc-2024}B. Fu, K.-Z. Bai, and S.-Q. Shen, Half-quantum
mirror Hall effect, Nat. Commun. \textbf{15}, 6939 (2024)

\bibitem[Wang(2021)]{Wang-prb-2021}H.-W. Wang, B. Fu, and S.-Q. Shen,
Helical symmetry breaking and quantum anomaly in massive Dirac fermions,
Phys. Rev. B \textbf{104}, L241111 (2021).

\bibitem[shen(2005)]{shen2005prb}S. Q. Shen, Y. J. Bao, M. Ma, X.
C. Xie, and F. C. Zhang, Resonant spin Hall conductance in quantum
Hall systems lacking bulk and structural inversion symmetry, Phys.
Rev. B \textbf{71}, 155316 (2005)

\bibitem[Wilczek(1987)]{Wilczek1987axiondynamics}F. Wilczek, Two
applications of axion electrodynamics, Phys. Rev. Lett. \textbf{58},
1799 (1987).

\bibitem[Schnyder(2008)]{Schnyder2008classification}A. P. Schnyder,
S. Ryu, A. Furusaki, and A. W. W. Ludwig,Classification of topological
insulators and superconductors in three spatial dimensions, Phys.
Rev. B \textbf{78}, 195125 (2008).

\bibitem[Ryu(2010)]{Ryu2010tenfold}S. Ryu, A. P. Schnyder, A. Furusaki,
and A. W. W. Ludwig, Topological insulators and superconductors: tenfold
way and dimensional hierarchy, New J. Phys. \textbf{12}, 065010 (2010).

\bibitem[Li(2010)]{Li2010np}R. Li, J. Wang, X. L. Qi, and S. C. Zhang,
Dynamical axion field in topological magnetic insulators. Nat. Phys.
\textbf{6}, 284(2010).

\bibitem[Essin(2009)]{Essin2009OMP}A. M. Essin, J. E. Moore, and
D. Vanderbilt, Magnetoelectric polarizability and axion electrodynamics
in crystalline insulators, Phys. Rev. Lett. \textbf{102}, 146805 (2009).

\bibitem[Spaldin(2005)]{Spaldin2005ME}N. A. Spaldin and M. Fiebig,
The renaissance of magnetoelectric multiferroics, Science \textbf{309},
391 (2005).

\bibitem[Fiebig(2005)]{Fiebig2005ME}M. Fiebig, Revival of the magnetoelectric
effect, J. Phys. D \textbf{38}, R123 (2005).

\bibitem[sekine(2014)]{sekine2014jpsj}A. Sekine and K. Nomura, Axionic
antiferromagnetic insulator phase in a correlated and spin--orbit
coupled System. J. Phys. Soc. Jpn. 83, 104709 (2014).

\bibitem[maciejko(2010)]{maciejko2010prl}J. Maciejko, X.-L. Qi, H.
D. Drew, and S.-C. Zhang, Topological Quantization in Units of the
Fine Structure Constant, Phys. Rev. Lett. \textbf{105}, 166803 (2010).

\bibitem[Tse(2011)]{Tse2011prb}W.-K. Tse and A. H. MacDonald, Magneto-optical
Faraday and Kerr effects in topological insulator films and in other
layered quantized Hall systems. Phys. Rev. B \textbf{84}, 205327(2011).

\bibitem[fu(2021)]{fu2021prr}B. Fu, Z. A. Hu, and S.Q. Shen, Bulk-hinge
correspondence and three-dimensional quantum anomalous Hall effect
in second-order topological insulators. Phys. Rev. Research \textbf{3},
033177(2021).

\bibitem[Lei(2024)]{Lei-2024-arxiv}C. Lei, P. T. Mahon, C. M. Canali,
and A. H. MacDonald, Capacitive detection of the topological magnetoelectric
effect, Phys. Rev. Lett. \textbf{133}, 246607 (2024).

\end{thebibliography}
\end{document}